\documentclass[10pt, twocolumn]{article}   

\usepackage{graphicx}
\usepackage{amsmath}
\usepackage{authblk}
\usepackage{abstract}
\usepackage{xkeyval}
\presetkeys{Gin}{width=0.494\textwidth}{} 
\usepackage[colorlinks=true,linkcolor=blue,citecolor=blue]{hyperref}
\usepackage[width=18.00cm, height=25.00cm]{geometry}
\usepackage{caption}
\usepackage[switch]{lineno}

\title{New Constraints on all flavour Galactic diffuse neutrino emission with the ANTARES telescope.}
\author[1]{A.~Albert}
\author[2]{M.~Andr\'e}
\author[3]{M.~Anghinolfi}
\author[4]{G.~Anton}
\author[5]{M.~Ardid}
\author[6]{J.-J.~Aubert}
\author[7]{T.~Avgitas}
\author[7]{B.~Baret}
\author[8]{J.~Barrios-Mart\'{\i}}
\author[9]{S.~Basa}
% ...

\affil[1]{\scriptsize{GRPHE - Universit\'e de Haute Alsace - Institut universitaire de technologie de Colmar, 34 rue du Grillenbreit BP 50568 - 68008 Colmar, France}}
\affil[2]{\scriptsize{Technical University of Catalonia, Laboratory of Applied Bioacoustics, Rambla Exposici\'o, 08800 Vilanova i la Geltr\'u, Barcelona, Spain}}
\affil[3]{\scriptsize{INFN - Sezione di Genova, Via Dodecaneso 33, 16146 Genova, Italy}}
\affil[4]{\scriptsize{Friedrich-Alexander-Universit\"at Erlangen-N\"urnberg, Erlangen Centre for Astroparticle Physics, Erwin-Rommel-Str. 1, 91058 Erlangen, Germany}}
\affil[5]{\scriptsize{Institut d'Investigaci\'o per a la Gesti\'o Integrada de les Zones Costaneres (IGIC) - Universitat Polit\`ecnica de Val\`encia. C/  Paranimf 1, 46730 Gandia, Spain}}
\affil[6]{\scriptsize{Aix Marseille Univ, CNRS/IN2P3, CPPM, Marseille, France}}
\affil[7]{\scriptsize{APC, Univ Paris Diderot, CNRS/IN2P3, CEA/Irfu, Obs de Paris, Sorbonne Paris Cit\'e, France}}
\affil[8]{\scriptsize{IFIC - Instituto de F\'isica Corpuscular (CSIC - Universitat de Val\`encia) c/ Catedr\'atico Jos\'e Beltr\'an, 2 E-46980 Paterna, Valencia, Spain}}
\affil[9]{\scriptsize{LAM - Laboratoire d'Astrophysique de Marseille, P\^ole de l'\'Etoile Site de Ch\^ateau-Gombert, rue Fr\'ed\'eric Joliot-Curie 38,  13388 Marseille Cedex 13, France}}
% ...

\date{}

\begin{document}

\onecolumn
\maketitle


\begin{onecolabstract}

        The flux of very high-energy neutrinos produced in our Galaxy by the interaction of accelerated cosmic rays with the interstellar medium is not yet determined. The characterization of this flux will shed light on Galactic accelerator features, gas distribution morphology and Galactic cosmic ray transport. The central Galactic plane can be the site of an enhanced neutrino production, thus leading to anisotropies in the extraterrestrial neutrino signal as measured by the IceCube Collaboration. The ANTARES neutrino telescope, located in the Mediterranean Sea, offers a favourable view on this part of the sky, thereby allowing for a contribution to the determination of this flux. The expected diffuse Galactic neutrino emission can be obtained linking a model of generation and propagation of cosmic rays with the morphology of the gas distribution in the Milky Way.
        In this paper, the so-called ``Gamma model'' introduced recently to explain the high-energy gamma ray diffuse Galactic emission, is assumed as reference. The neutrino flux predicted by the ``Gamma model'' depends of the assumed primary cosmic ray spectrum cut-off. Considering a radially-dependent diffusion coefficient, this proposed scenario is able to account for the local cosmic ray measurements, as well as for the Galactic gamma ray observations. Nine years of ANTARES data are used in this work to search for a possible Galactic contribution according to this scenario. All flavour neutrino interactions are considered. No excess of events is observed and an upper limit is set on the neutrino flux of $1.1$ ($1.2$) times the prediction of the ``Gamma model'' assuming the primary cosmic ray spectrum cut-off at 5 (50) PeV.
         This limit excludes the diffuse Galactic neutrino emission as the major cause of the ``spectral anomaly'' between the two hemispheres measured by IceCube.

\end{onecolabstract}
\clearpage

\twocolumn[]
\section{Introduction}
        
        The Fermi-LAT telescope obtained detailed measurements of diffuse high-energy gamma ray emission along the Galactic plane after the subtraction of point-like contributions~\cite{2012ApJ...750....3A}. Above a few GeV most of this observed diffuse emission can be attributed to photons produced in neutral pion decays coming from primary cosmic ray (CR) interactions with the ambient medium (dust, molecular clouds, etc). In these hadronic processes, a neutrino counterpart emission is also expected from $\pi^{+/-}$ decays. The good coverage of the Southern Hemisphere, as well as its large effective area and good angular resolution, allows the ANTARES neutrino telescope to probe models for this expected flux.

        Detailed computations of the neutrino flux produced in this context have been carried out. Using radially-dependent CR diffusion properties, a novel comprehensive interpretation of CR transport in our galaxy was used.
        With this model the observed local CR features~\cite{2013APh....47...54A,2011Sci...332...69A,2015PhRvL.114q1103A}, as well as the diffuse Galactic gamma ray emission measured by Fermi-LAT~\cite{2012ApJ...750....3A}, H.E.S.S.~\cite{2006Natur.439..695A} and Milagro~\cite{2008ApJ...688.1078A}, can be reproduced.
        The new model, developed under this scenario, called ``KRA$_\gamma$'' or ``Gamma model''~\cite{2015ApJ...815L..25G,Gaggero:2014xla,2017arXiv170201124G} is used in this paper. It allows the prediction of the expected full sky neutrino flux induced by Galactic CR interactions. Compared to conventional scenarios where a homogeneous CR transport is assumed for the whole Galactic plane~\cite{2016PhRvD..93a3009A}, an enhanced neutrino emission up to five times larger in its central part is predicted~\cite{2015arXiv150803681G}. 
        The spectrum of primary interacting CRs of the ``Gamma model'' presents a hardening around 250 GeV per nucleon as observed by PAMELA~\cite{2011Sci...332...69A} and AMS-02~\cite{2015PhRvL.114q1103A} experiments. Above this energy, the CR source spectra extend steadily up to an exponential cut-off on the energy per nucleon E$_{\text{cut}}$. 
        Two representative values of this quantity have been considered,  namely E$_{\text{cut}}$  = 5 and 50 PeV, which -- for the ``Gamma model'' setup -- match CREAM proton and helium data~\cite{2010ApJ...714L..89A} and roughly reproduce KASCADE~\cite{KASCADE2005} and KASCADE Grande data~\cite{2013APh....47...54A}. While the KASCADE proton data favor the lowest cut-off (5 PeV), the highest one (50 PeV) is favored by the KASCADE-Grande all-particle spectrum.

        The two different cut-off cases of the ``Gamma model'' will be referred to as the two ``reference models'' in this article.
        The morphological and energetic characteristics of the neutrino fluxes computed from these models are obtained by linking the DRAGON code~\cite{Dragonweb} for Galactic CR transport, using the gas 3D distribution described in Ref.~\cite{Ferriere2007} for Galactocentric radii $R < 1.5$ kpc, and the gas ring model used by the Fermi collaboration~\cite{2012ApJ...750....3A} for larger radii.

        In the last few years, the IceCube Collaboration has reported a significant excess of high-energy neutrinos with respect to the expected atmospheric background~\cite{2013PhRvL.111b1103A,2015arXiv151005223T,2014ApJ...796..109A}.
        The spectral energy distribution obtained with 4 years of ``high-energy starting events'' (HESE) through a full sky analysis results in a one flavour normalisation factor $E^{2}\Phi(E)=2.2 (\pm0.7) \cdot (E/ 100 \text{ TeV})^{-0.58} \times10^{-8}$ GeV cm$^{-2}$ s$^{-1}$ sr$^{-1}$ with a fitted spectral index $\alpha=2.58\pm0.25$~\cite{2015arXiv151005223T}.
        Nevertheless, a dedicated analysis with 6 years of muonic neutrinos from the Northern Hemisphere shows a normalisation factor of $E^{2}\Phi(E)=0.90^{+0.3}_{-0.27} \cdot (E/ 100 \text{ TeV})^{-0.13} \times10^{-8}$ GeV cm$^{-2}$ s$^{-1}$ sr$^{-1}$ and  a spectral index $\alpha=2.13\pm0.13$~\cite{2016ApJ...833....3A} generating a non-negligible discrepancy between the measured neutrino spectral energy distributions of the two hemispheres, the so-called ``spectral anomaly''.

        Different explanations have been put forth for the tension in the normalisation versus spectral index between the two contributions leading to a relative enhanced emission in the Southern Hemisphere. One of them is a cut-off in the spectrum as these two analyses have a different energy threshold.
        Another one comes from the position of the Milky Way. As its central region is at negative declinations, the sum of a Galactic and an extragalactic component~\cite{2016JCAP...12..045P, 2016EPJWC.11604009M} can result in different spectral behaviours in the two hemispheres.
        From a statistical point of view, $50\%$ of the observed IceCube cosmic neutrino signal events are compatible with a Galactic plane origin~\cite{2016APh....75...60N}.
        Conversely, when considering the reference model with 50 PeV cut-off, it is possible to account for a maximum of $18\%$ of the full sky HESE flux measured by IceCube, while in the conventional scenario, only $8\%$ of this flux can be related to Galactic diffuse emission~\cite{2015ApJ...815L..25G}.

        The ANTARES view of the Southern Sky, its exposure towards the Galactic centre region and its very  good angular resolution makes it well suited to either detect the neutrino flux predicted by the reference models over several decades in energy, or place competitive upper limits on the flux normalisation. In order to fully exploit the particular morphology of the expected signal, as well as the angular dependency of the energy spectrum, a maximum likelihood analysis is performed assuming the signal events have the angular and energy distributions obtained from the reference models.
        With this technique, a new stringent upper limit is obtained on the neutrino flux over three decades in energy based on 9 years of data taking.

        The paper is structured as follows: A description of the detector and the dataset is provided, followed by a description of the maximum likelihood analysis and then a discussion on the results. A very important consequence of this paper is the strong disfavour of the diffuse Galactic emission as the origin of the spectral anomaly observed by IceCube. 
        \newline

\section{The ANTARES detector and data sample}
        The ANTARES neutrino telescope~\cite{Antares11a} is installed at 2475~m depth in the Mediterranean sea, 40~km off the coast of Toulon, France. It is made of an array of photomultipliers, which detect Cherenkov light induced by particles created during high energy neutrino interactions. Two detection channels are available for neutrinos above a few tens of GeV: charged current interactions of muon neutrinos, with the subsequent Cherenkov emission by the outgoing muon, which constitute most of the so-called ``track events''. All other interactions, which produce electromagnetic or hadronic showers in the detector, representing the so-called ``shower events''. For the former, the sub-degree angular resolution and an energy accuracy of the order of a fraction of a decade can be obtained; they benefit from the kilometer-scale muon track length to enlarge the effective detection volume thereby increasing the event rates. The latter type of events has an angular accuracy of a few degrees, but an energy resolution of~10\%; these performances are achievable only in a smaller effective detection volume, thus reducing the neutrino effective area.

        A Monte Carlo simulation of electron and muon neutrinos and antineutrinos has been used in this analysis. The contribution of tau neutrinos has been estimated by scaling up in a consistent manner the number of electron and muon neutrinos.
        The data used in this search was recorded between the 29$^{\rm th}$ of January 2007 and the 31$^{\rm st}$ of December 2015 for a total livetime of 2423.6 days. Monte Carlo simulations reproduce the time variability of the detector conditions according to a ``run-by-run'' approach~\cite{fusco_run-by-run_2016}.

        The background consists of atmospheric neutrinos and downward-going muons created by CR-induced atmospheric air showers. While atmospheric neutrinos cannot be distinguished on an event-by-event basis from cosmic neutrinos, the event selection aims at suppressing events from downward-going muons by selecting events reconstructed as upward-going. This procedure follows the same steps as the one used for the search of point-like sources in Ref.~\cite{michael_neutrino_2015}. The selection of events in this analysis maximizes the discovery power (defined in section~\ref{sec:search_method}) of the flux predicted by the reference model with the 50 PeV cut-off when using the search method described below.
        
        An event is selected as track-like if it is reconstructed by the tracking algorithm~\cite{noauthor_first_2013} as upward-going and if it passes the selection cuts defined in the searches for point-like neutrino sources~\cite{Albert:2017ohr}. This rejects most of the background from CR-induced atmospheric muons. Shower-like events are selected if they are not present in the track sample and if the event is reconstructed within a fiducial volume surrounding the apparatus with high quality by the shower reconstruction algorithm~\cite{michael_neutrino_2015}. These events must also be reconstructed as upward-going. The dataset consists of 7300 tracks and 208 showers events. The median angular resolution for tracks and showers is $0.6^\circ$ and $2.7^\circ$ respectively, when considering the reference model with the 5 PeV cut-off. For the reference model with the 50 PeV cut-off, the median angular resolution for tracks improves to $0.5^\circ$ whereas the one for showers does not change significantly.
        \newline

        \section{Search Method}\label{sec:search_method}
        The analysis presented in this work is based on a likelihood ratio test, widely used in neutrino astronomy, e.g. in the search for neutrinos from individual point-like or extended sources by ANTARES~\cite{Tino_ICRC, PS_Combined_first_2016, GRB_stacked_2017, x-binaries}.
        It is adapted here to a full-sky search where the signal map is built according to the reference models mentioned above. A probability density function of observables was defined according to given expectations/models. Data are considered to be a mixture of signal and background events, so the likelihood function is defined as:

        \begin{equation}
        \begin{split}
                \mathcal{L}_\text{sig+bkg} = \prod\limits_{\mathcal{T} \in \{\text{tr}, \text{sh}\}} \prod\limits_{i \in \mathcal{T}}& [\mu_\text{sig}^\mathcal{T}\cdot pdf_\text{sig}^\mathcal{T}(E_i, \alpha_i, \delta_i)\\
                & + \mu_\text{bkg}^\mathcal{T}\cdot pdf_\text{bkg}^\mathcal{T}(E_i, \theta_i, \delta_i)]
        \end{split}
        \end{equation}

        \noindent
        where $E_i$ is the reconstructed energy, $\alpha_i$ and $\delta_i$ the right ascension and declination (equatorial coordinates), and $\theta_i$ the zenith angle of the event $i$. For each event topology $\mathcal{T}$ (track or shower), given a total number of events $\mu_\text{tot}^\mathcal{T}$, the number of background events $\mu_\text{bkg}^\mathcal{T}$ corresponds to $\mu_\text{tot}^\mathcal{T} - \mu_\text{sig}^\mathcal{T}$.
        The number of signal events $\mu_\text{sig}^\mathcal{T}$ is fitted by maximising the likelihood, allowing only non-negative values.
        The signal and background probability density functions of an event are defined as:
        \begin{equation}
                pdf_\text{sig}^\mathcal{T}(E_i, \alpha_i, \delta_i) = \mathcal{M}_\text{sig}^\mathcal{T}(\alpha_i, \delta_i) \cdot \mathcal{E}_\text{sig}^\mathcal{T}(E_i, \alpha_i, \delta_i)
        \end{equation}
        \begin{equation}
                pdf_\text{bkg}^\mathcal{T}(E_i, \theta_i, \delta_i) = \mathcal{M}_\text{bkg}^\mathcal{T}(\delta_i) \cdot \mathcal{E}_\text{bkg}^\mathcal{T}(E_i, \theta_i)
        \end{equation}

        \begin{figure}
                \centering
                \includegraphics {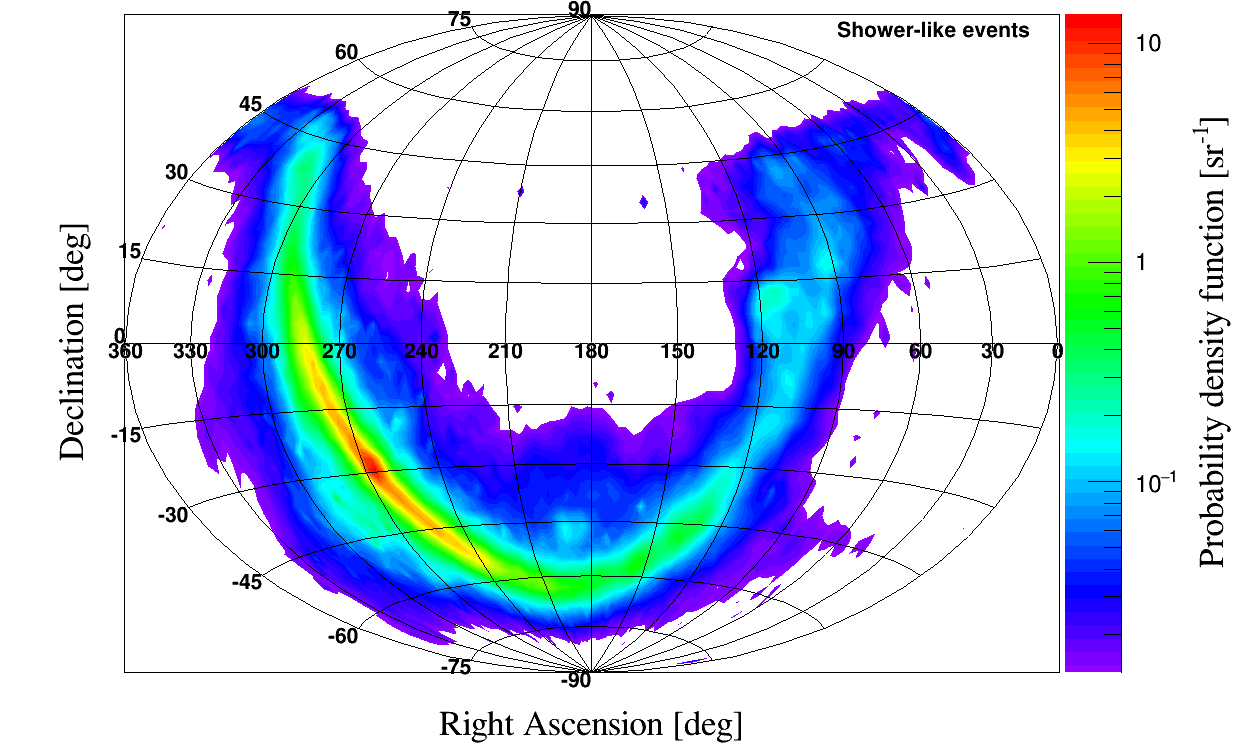}
                \includegraphics {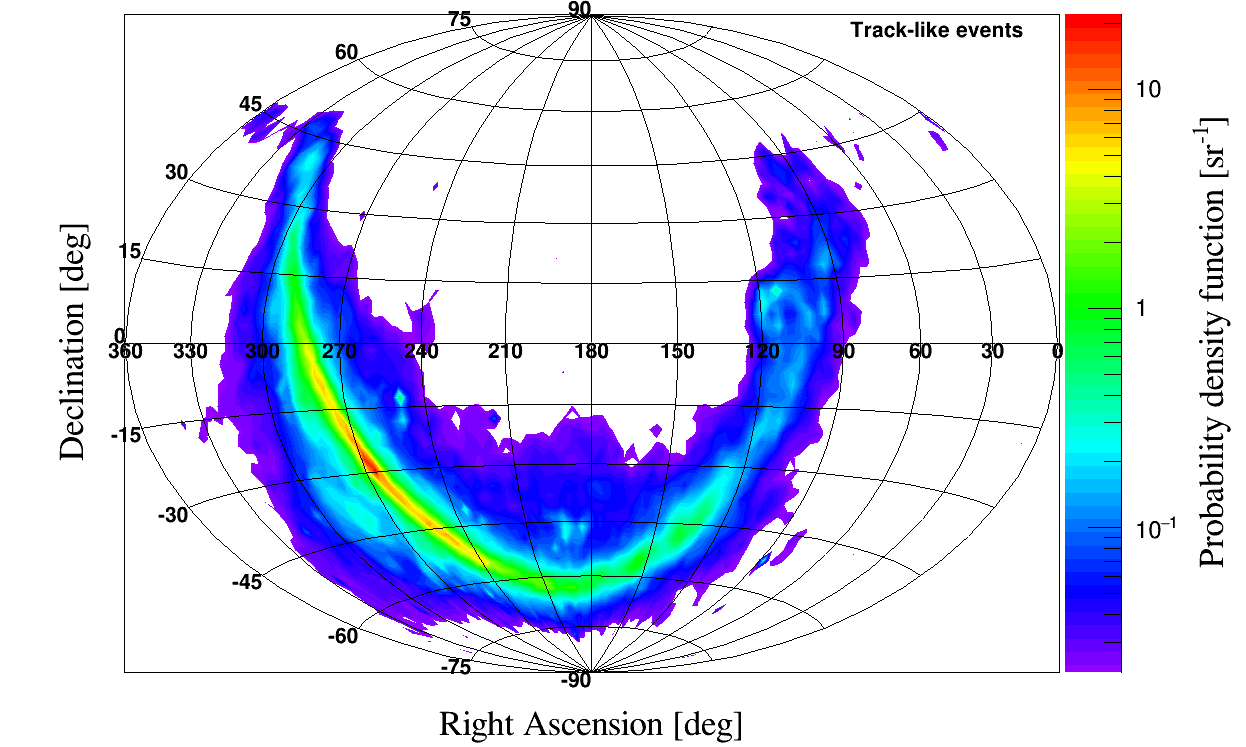}
                \caption{Probability density function of the reconstructed direction of signal events $\mathcal{M}_\text{sig}^\mathcal{T}$, in equatorial coordinates for shower-like (top) and track-like (bottom) events.}\label{fig:Msig}
        \end{figure}

        \noindent
        where $\mathcal{M}^\mathcal{T}$ are the probability density functions to reconstruct an event in a given position in the sky. The probability density functions $\mathcal{M}_\text{sig}^\mathcal{T}$, shown in Figure~\ref{fig:Msig} (for the 5 PeV energy cut-off model) as obtained from Monte Carlo simulation, depend on the differential neutrino fluxes predicted by the reference models folded with the detector response to a given direction in the sky. The background distribution $\mathcal{M}_\text{bkg}^\mathcal{T}$ is obtained from the data, by scrambling the time of events which results in a randomisation of the corresponding right ascension. This is a conservative estimate of the background. Moreover, provided that the signal is weak enough (which is the case given the non detection of a diffuse flux from the Galactic ridge~\cite{on-off}), this procedure produces a background distribution which only depends on the declination. This is due to the fact that Earth's rotation and uniform distribution of the time the detector was operational imply a flat atmospheric background right ascension distribution.
        The parameter $\mathcal{E}^\mathcal{T}$ is the probability density function of the reconstructed energy. For the signal, $\mathcal{E}_\text{sig}^\mathcal{T}$ depends on the equatorial coordinates as the energy spectra of the reference models depend on the position in the sky. The parameter $\mathcal{E}_\text{bkg}^\mathcal{T}$ depends on the corresponding local zenith $\theta_i$ to account for potential reconstruction systematic effects due to the detector response.

        The test statistic $\mathcal{Q}$ is then defined as the logarithm of the likelihood ratio:

        \begin{equation}
                \mathcal{Q} = \log_{10}(\mathcal{L}_\text{sig+bkg}) - \log_{10}(\mathcal{L}_\text{bkg})
        \end{equation}

        \noindent
        with $\mathcal{L}_\text{bkg} = \mathcal{L}_\text{sig+bkg}(\mu_\text{sig}^\text{sh}=\mu_\text{sig}^\text{tr}=0)$.

        The discovery power and sensitivity of the search are computed by building the probability density functions of the test statistic $pdf_\Phi(\mathcal{Q})$ assuming different values of the normalisation factor $\Phi$ of the reference model fluxes. The discovery power is defined as the probability for a given signal normalisation to yield a test statistic value corresponding to a $3\sigma$ significance excess on top of the expected background. For a given value of the test statistic $\mathcal{Q}_{obs}$ compatible with background expectation, the $90\%$ upper limit will be defined as the highest signal normalisation which would yield a test statistic value above $\mathcal{Q}_{obs}$ $90\%$ of the time. The sensitivity of the search is then defined as the average of the upper limits corresponding to all possible $\mathcal{Q}$ values in the background hypothesis ($\Phi = 0$) weighted by their probabilities $pdf_{\Phi=0}(\mathcal{Q})$.  Pseudo-experiments are thus produced, varying the number of signal events $\mu_\text{sig}^\text{sh+tr}$ accordingly.
        They are generated using the probability density functions $\mathcal{M}^\mathcal{T}$ and $\mathcal{E}^\mathcal{T}$ defined above. A total of $10^5$ pseudo-experiments are produced in the background case ($\mu_\text{sig}^\text{sh+tr} = 0$) and $10^4$ for each value of $\mu_\text{sig}^\text{sh+tr}$ in the range [1,55] where the rate of showers, taken from the Monte Carlo simulation, is $\sim$20\% of $\mu_\text{sig}^\text{sh+tr}$.
		For each pseudo-experiment, the number of fitted track ($\mu_\text{fit}^\text{tr}$) and shower ($\mu_\text{fit}^\text{sh}$) events can be obtained.
        
        The distribution of $[\mu_\text{sig}^\text{sh+tr}-(\mu_\text{fit}^\text{tr}+\mu_\text{fit}^\text{sh})]$ has a median value close to zero and a standard deviation $\sigma^* = 13$ for the model with the 5 PeV cut-off and $\sigma^* = 12$ with the 50 PeV cut-off. It is worth noticing that the value of $\sigma^*$ is related to the background fluctuation, which does not change when varying the true number of signal events for a given model. This means that, if the exposure increases by a given factor, $\sigma^*$ increases less rapidly.
        The probability density functions of $\mathcal{Q}$ for integer numbers of signal events $pdf_{\mu_\text{sig}^\text{sh+tr}}(\mathcal{Q})$ are obtained from pseudo-experiments. They are linked to $pdf_\Phi(\mathcal{Q})$, with $\Phi$ leading to a mean number of detected signal events $n$, by:

        \begin{equation}
                pdf_\Phi(\mathcal{Q}) = \sum\limits_{\mu_\text{sig}^\text{sh+tr}} P(\mu_\text{sig}^\text{sh+tr}|n)\cdot pdf_{\mu_\text{sig}^\text{sh+tr}}(\mathcal{Q})\label{pdf_flux}
        \end{equation}

        \noindent
        where $P$ is the Poissonian probability distribution.

        The systematic uncertainty on the acceptance of the ANTARES photomultipliers implies an uncertainty of 15\% on the effective area~\cite{AdrianMartinez:2012rp}. To account for this, the number of expected signal events $n$ from a given flux is fluctuated using a Gaussian distribution with a standard deviation of 15\%. An uncertainty on the background distribution due to statistical fluctuations in the data is also taken into account by fluctuating $\mathcal{M}_\text{bkg}^\mathcal{T}(\delta_i)$.

        \begin{figure}
                \includegraphics{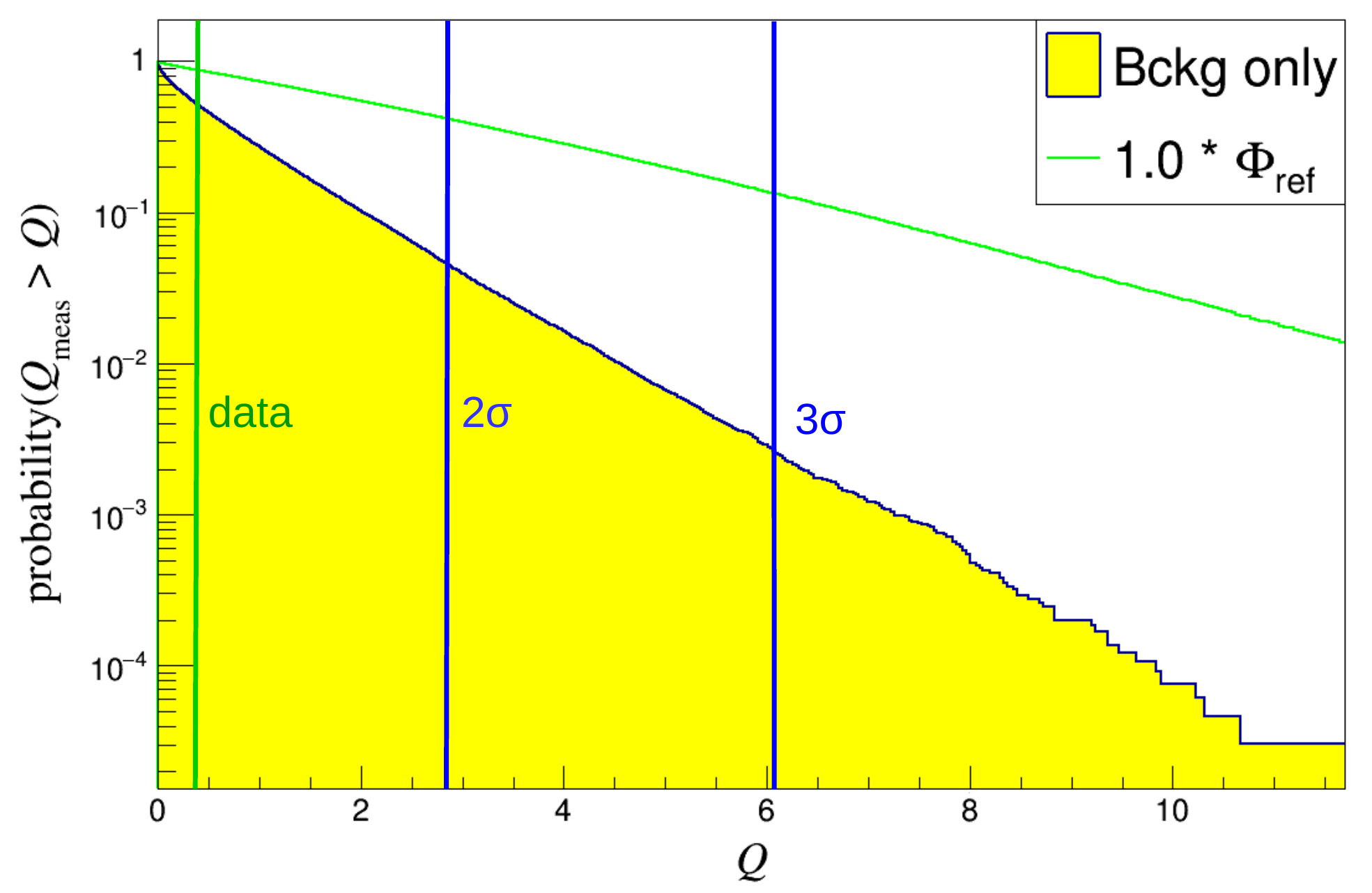}
                \caption{Anti-cumulative distribution of the test statistic $\mathcal{Q}$ from the pseudo-experiments for background-only (yellow area) and with signal from the reference model with the 5 PeV cut-off (red line). The corresponding values of the test statistic for $2\sigma$ and $3\sigma$ confidence level are shown (blue lines) along with the value from data (green line).}\label{fig:cdfTS}
        \end{figure}

        The p-value for a given $\mathcal{Q}$ is defined as the probability to measure a test statistic larger than this one in the background-only case. It is given by the anti-cumulative probability density function of $\mathcal{Q}$ with no injected signal (Figure~\ref{fig:cdfTS}). Upper limits at a given confidence level are set according to the corresponding distributions with injected signal events.

        \begin{figure}
            \includegraphics{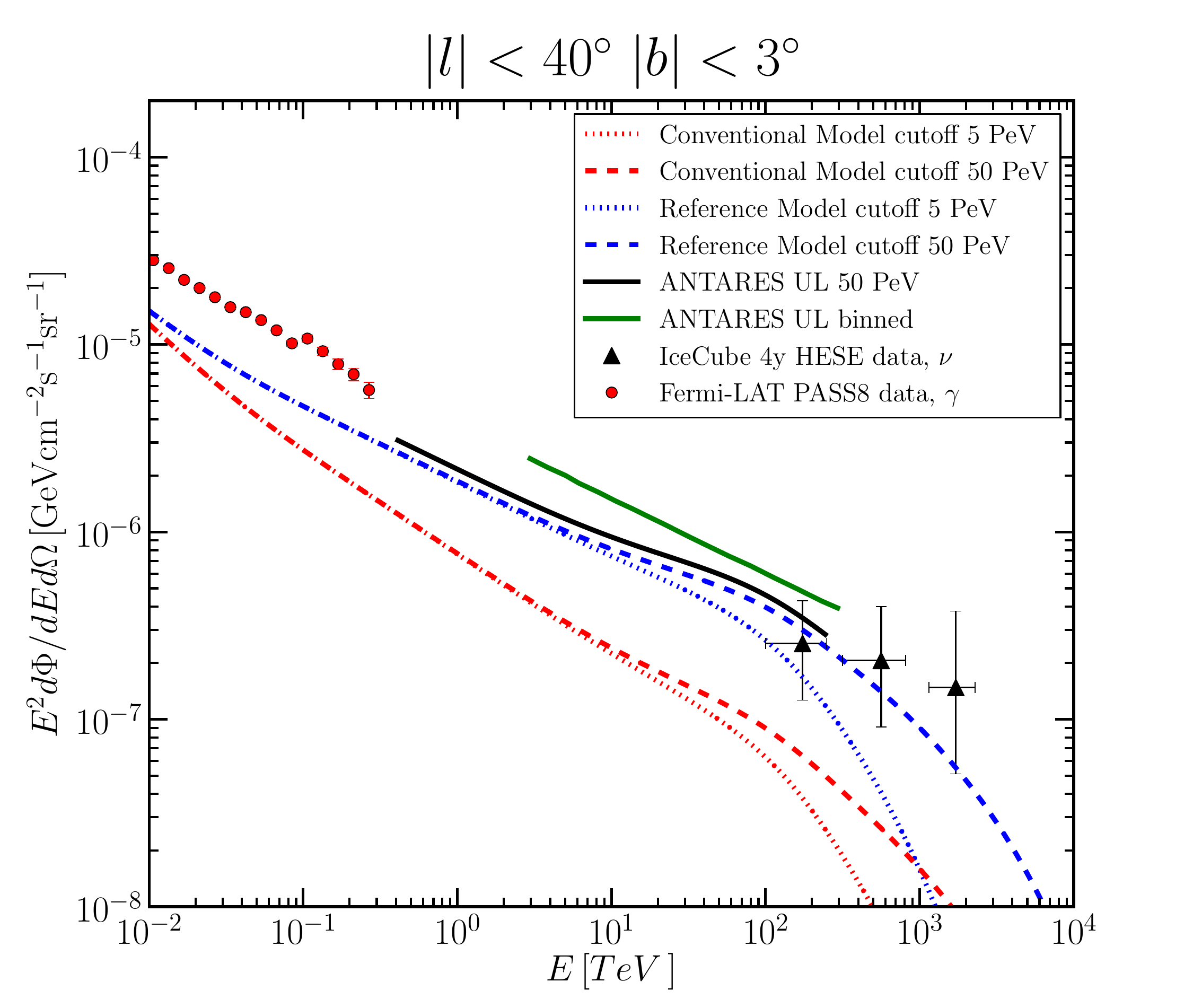}
                \caption{ANTARES upper limit at 90\% confidence level on the three flavour neutrino flux (solid black line) on the reference model with a 50 PeV energy cut-off (blue dashed line). The neutrino fluxes according to the reference model with the 5 PeV energy cut-off (blue dotted line), the conventional model with the 50 PeV (red dashed line) and 5 PeV (red dotted line) cut-offs are shown for all neutrino flavours, as well as the previously published ANTARES upper limit~\cite{on-off} (solid green line) and the 4 years of HESE reconstructed by IceCube (black triangles). The diffuse gamma ray spectral energy distribution derived from PASS8 Fermi-LAT data (red points) is also presented here. These expectations and results concern the inner Galactic plane region ($|l| < 40^\circ$ and $|b| < 3^\circ$).}\label{fig:limit}
        \end{figure}

        For the model with the 5 PeV cut-off, 90\% of signal events are in the energy range [$3.5 \cdot 10^{-1}$,$1.3 \cdot 10^{2}$] TeV for track-like events and between [$2.0$,$1.5 \cdot 10^{2}$] TeV for shower-like events. For the 50 PeV cut-off, these energy ranges are [$4.0 \cdot 10^{-1}$,$2.3 \cdot 10^{2}$] TeV for the tracks and [$2.2$,$2.6 \cdot 10^{2}$] TeV for the showers.
        To avoid biasing the analysis, the data have been blinded by time-scrambling. Both the sensitivity and the discovery power of the analysis are derived from this blinded dataset. The sensitivity, defined as the average upper limit at 90$\%$ confidence level, is $1.4 \times \Phi_\text{ref}^5$ when a cut-off for CR primary protons at $5$ PeV is set. A mean of $\mu^* = 11.6$ signal events is expected from the model. It corresponds to the sum of track-like and shower-like events, with showers representing $\sim$20\% of the total.
		The resulting discovery power at 3$\sigma$ confidence level is $7\%$. For the model with a 50 PeV cut-off, the sensitivity is $1.05 \times \Phi_\text{ref}^{50}$ and $\mu^* = 13.7$ signal events are expected, resulting in a discovery power of $14\%$ for a 3$\sigma$ confidence level.
        \newline

\section{Results}
        After unblinding, the test statistic of the data is computed. The corresponding $\mathcal{Q}$ value is shown as the green line in Figure~\ref{fig:cdfTS}.
        Table~\ref{table:results} presents the results for the two different cut-off energies (column~1) considered by the models.
        Column~2 reports the number of expected events, $\mu^*$, and column~3 the standard deviation of the distribution of the number of fitted events, $\sigma^*$, which are defined in section~\ref{sec:search_method}.

        For the data sample, the numbers of fitted track-like events, $\mu_\text{data}^\text{tr}$, and shower-like events, $\mu_\text{data}^\text{sh}$, are reported in columns~4 and 5, respectively. Their sum is smaller than $\mu^*$, but still compatible with the expected fluctuations. These include the Gaussian fluctuations. due to the background (which is within $1 \sigma^*$) and the Poissonian fluctuations on the number of signal events.

        Finally, using the anti-cumulative distribution of the background test statistic, the p-value of the data -- as defined in section~\ref{sec:search_method} -- is computed and reported in column~6.
        The derived upper limits at 90\% confidence level on the reference models are reported in the last column of Table~\ref{table:results}.

        \begin{table*}
        \begin{center}
        \caption{Results of the presented analysis for the two reference models corresponding to different energy cut-offs. The number of expected signal events, $\mu^*$, is shown, as well as $\sigma^*$, the standard deviation of the distribution of the difference between the number of fitted events and the number of injected events in the pseudo-experiment. For the data sample, the numbers of fitted shower-like events, $\mu_\text{data}^\text{sh}$, and track-like events, $\mu_\text{data}^\text{tr}$, are reported with the p-values and the upper limits at 90\% confidence level.\label{table:results}}
        \begin{tabular}{ c  c  c  c  c  c  c }
                \hline
            \hline
                Energy cut-off & $\mu^*$ & $\sigma^*$ & $\mu_\text{data}^\text{sh}$ & $\mu_\text{data}^\text{tr}$ & p-value & UL at 90\% CL \\
                \hline
            5 PeV & 11.6 & $13$ & $1.9$ & $2 \times 10^{-3}$ & $0.67$ & $1.1 \times \Phi_\text{ref}^5$\\
            50 PeV & 13.7 & $12$ & $2.6$ & $7 \times 10^{-4}$ & $0.54$ & $1.2 \times \Phi_\text{ref}^{50}$\\
                \hline
                \hline
        \end{tabular}
        \end{center}
        \end{table*}

        Figure~\ref{fig:limit} shows the 90\% confidence level upper limit of this analysis that relies on the particular morphology and energy spectrum of the reference model. The dotted blue line refers to the reference model assuming a cut-off of 5 PeV for the primary protons, which produce neutrinos when interacting with gas. Although full sky data were used in this analysis, the expectations and the results concerning the inner Galactic plane region ($|l| < 40^\circ$ and $|b| < 3^\circ$) are shown on this plot. This allows the presented limit and the previous ANTARES constraint on the neutrino emission~\cite{on-off} from the same region to be compared. The diffuse gamma ray spectral energy distribution derived from PASS8 Fermi-LAT data~\cite{atwood_pass_2013} obtained after the subtraction of point-like components comprised in this region is also shown for comparison.
        And the red dashed line shows the predicted spectrum from the conventional model with homogeneous CR diffusion. The neutrino flux from the 4 year IceCube HESE catalog for individual events with origin compatible with this region is shown as black triangles. All flavour neutrino fluxes are represented in this figure.
        \newline

\section{Conclusions}
        The study reported here is based on nine years of ANTARES data collected from 2007 to 2015. It uses a likelihood ratio test to search for a diffuse Galactic-dominated neutrino flux, characterised by the recently introduced ``Gamma model'' used as reference model. As a result, a neutrino flux with normalisation factor of $1.1 \times \Phi_\text{ref}^5$ (resp. $1.2 \times \Phi_\text{ref}^{50}$) is excluded at 90\% confidence level when the model with the 5 PeV cut-off (resp. 50 PeV) is considered.

        Using neutrinos of all flavours as well as a larger amount of data leads to an improvement in the sensitivity and more stringent upper limits with respect to the previous ANTARES analysis~\cite{on-off}.  The new upper limits do not extend above $\sim$200 TeV due to the significant softening of the spectrum. The additional gain in sensitivity below 3 TeV with respect to the previous analysis results from the usage of a new unbinned  method that uses spatial and energy information. At low energies, the limit obtained from this analysis reaches almost the high-energy tail of the Fermi-LAT sensitivity.

        Noticing the enhanced Galactic hadronic emission predicted by the reference models with respect to a conventional scenario, the obtained limits represent a strong constraint on a possible diffuse neutrino emission from the Galactic plane.

        Considering the flux upper limit with 90\% confidence level shown in Table~\ref{table:results} for the 50 PeV cut-off, at most 18\% of the cosmic neutrino events measured by IceCube with the HESE dataset can originate from diffuse Galactic CR interaction. This corresponds to about 5.2 out of the 28.6 HESE with energy above 60 TeV expected to be cosmic neutrinos, as reported in Ref.~\cite{Aartsen:2014cva}. This limit is more restrictive than that allowed in Ref.~\cite{2016JCAP...12..045P, 2016APh....75...60N}.
        The reference model produces a larger North/South asymmetry than the conventional scenario: more than $\sim$80\% of the events are expected from the Southern hemisphere.
        Nevertheless, the contribution of the diffuse Galactic component to the difference between the observed number of HESE arising from the two hemispheres cannot be larger than 3.3 HESE, i.e. $\sim$10\% of the full sky flux. As a result, the neutrino flux produced by the Galactic CR interaction with gas cannot explain by itself the IceCube spectral anomaly. 
        These considerations are even more restrictive for the case of the 90\% confidence level upper limit corresponding to a primary CR cut-off of 5 PeV, as evident from the predicted flux given in Figure~\ref{fig:limit}.
        \newline

\section{Acknowledgements}
The authors acknowledge the financial support of the funding agencies:
Centre National de la Recherche Scientifique (CNRS), Commissariat \`a
l'\'ener\-gie atomique et aux \'energies alternatives (CEA),
Commission Europ\'eenne (FEDER fund and Marie Curie Program),
Institut Universitaire de France (IUF), IdEx program and UnivEarthS
Labex program at Sorbonne Paris Cit\'e (ANR-10-LABX-0023 and
ANR-11-IDEX-0005-02), Labex OCEVU (ANR-11-LABX-0060) and the
A*MIDEX project (ANR-11-IDEX-0001-02),
R\'egion \^Ile-de-France (DIM-ACAV), R\'egion
Alsace (contrat CPER), R\'egion Provence-Alpes-C\^ote d'Azur,
D\'e\-par\-tement du Var and Ville de La
Seyne-sur-Mer, France;
Bundesministerium f\"ur Bildung und Forschung
(BMBF), Germany;
Istituto Nazionale di Fisica Nucleare (INFN), Italy;
Stichting voor Fundamenteel Onderzoek der Materie (FOM), Nederlandse
organisatie voor Wetenschappelijk Onderzoek (NWO), the Netherlands;
Council of the President of the Russian Federation for young
scientists and leading scientific schools supporting grants, Russia;
National Authority for Scientific Research (ANCS), Romania;
Mi\-nis\-te\-rio de Econom\'{\i}a y Competitividad (MINECO):
Plan Estatal de Investigaci\'{o}n (refs. FPA2015-65150-C3-1-P, -2-P and -3-P, (MINECO/FEDER)), Severo Ochoa Centre of Excellence and MultiDark Consolider (MINECO), and Prometeo and Grisol\'{i}a programs (Generalitat
Valenciana), Spain;
Ministry of Higher Education, Scientific Research and Professional Training, Morocco.
We also acknowledge the technical support of Ifremer, AIM and Foselev Marine
for the sea operation and the CC-IN2P3 for the computing facilities.

\bibliographystyle{unsrt}
\bibliography{biblio}

\end{document}